\documentclass[preprint]{revtex4}%
\usepackage{graphicx}
\usepackage{dcolumn}
\usepackage{bm}
\usepackage{amsmath}
\usepackage{amsfonts}
\usepackage{amssymb}%
\begin{document}
\title{{\it Ab-initio} procedure for effective models of 
correlated materials with entangled band structure}
\author{Takashi Miyake$^{1,4}$} 
\author{Ferdi Aryasetiawan$^{2,4}$}
\author{Masatoshi Imada$^{3,4}$}
\affiliation{$^1$Research Institute for Computational Sciences, AIST, Tsukuba 305-8568, Japan}
\affiliation{$^2$Graduate School of Advanced Integration Science, Chiba University, Chiba, 263-8522 Japan}
\affiliation{$^3$Department of Applied Physics, University of Tokyo, Hongo, Tokyo 113-8656, Japan}
\affiliation{$^4$Japan Science and Technology Agency, CREST, Kawaguchi 332-0012, Japan}
\date{\today}

\begin{abstract}
We present a first-principles method for deriving 
effective low-energy models of electrons
in solids having entangled band structure. 
The procedure starts with dividing the Hilbert space into two subspaces, 
the low-energy part (``$d$ space'') and the rest of the space (``$r$ space''). 
The low-energy model is constructed for the $d$ space by eliminating the 
degrees of freedom of the $r$ space. 
The thus derived model contains 
the strength of electron correlation expressed by 
a partially screened Coulomb interaction, calculated 
in the constrained random-phase-approximation (cRPA) 
where screening channels within the $d$ space, $P_d$, are subtracted. 
One conceptual problem of this established downfolding method is that 
for entangled bands it is not clear how to cut out the $d$ space and 
how to distinguish $P_d$ from the total polarization.
Here, we propose a simple procedure to overcome this difficulty. 
In our scheme, the $d$ subspace is cut out from the Hilbert space of 
the Kohn Sham eigenfunctions with the help of a procedure 
to construct a localized Wannier basis. 
The $r$ subspace is constructed as the complementary space orthogonal 
to the $d$ subspace. After this disentanglement, $P_d$ becomes well defined. 
Using the disentangled bands, 
the effective parameters are uniquely determined in the cRPA. 
The method is successfully applied to 3$d$ transition metals. 
\end{abstract}

\maketitle

\section{Introduction}

In the last several decades many new materials with intriguing properties
were discovered and synthesized. These materials range from the
high-temperature superconductors to magnetic materials, and the latter have 
already found
real applications in electronic industry. Typically, most of these materials
contain elements from the 3$d$ or 4$f$ rows and their electronic structure is
characterized by the presence of a partially filled narrow band across the
Fermi level. The fact that the narrow band is partially filled implies that
there are many configurations with approximately equal weight rendering a
one-particle description of the electronic structure problematic. Indeed, it
has been recognized for a long time that many of the intriguing properties of
these materials originate from correlations among the electrons residing in
the partially filled narrow band. The electrons are neither fully localized,
like core electrons, nor itinerant, like $s$ or $p$ electrons in alkalis or
conventional semiconductors such as silicon or diamond. This hybrid property
poses a tremendous theoretical difficulty for an accurate description of the
electronic structure, because due to the electrons' partially itinerant
character the problem can neither be treated as a purely atomic problem 
nor within a pure band picture. Moreover,
the interaction with other electrons can be very important.

A large amount of work has been directed to solving the correlation problem
of the above materials. The usual approach is to consider only the narrow bands
near the Fermi level and eliminate the degrees of freedom for 
the rest of the bands by the downfolding procedure, 
resulting in the well-known Hubbard model
which contains an effective on-site Coulomb interaction, the Hubbard $U$. 
In general, the models represent multi-band systems containing interorbital 
as well as long-ranged part of Coulomb interaction. 
The models can then be solved with various sophisticated 
low-energy solvers such as dynamical mean-field theory (DMFT) \cite{georges96}
or solvers for lattice models \cite{imada}. 

An important issue in mapping the real system to a model Hamiltonian is how to
determine the one-particle kinetic energy term and 
the effective interaction or the Hubbard \emph{U} in the model. 
Unlike the
one-particle parameters that can be downfolded from the band structure, the
Hubbard \emph{U} is much more elusive to determine and it is often treated as
an adjustable parameter. A widely used scheme to calculate the Hubbard
\emph{U} from first-principles is the constrained LDA (cLDA) method
\cite{gunnarsson89,gunnarsson90,anisimov91}. The cLDA method, however, is
known from early on to yield values of $U$, which are too large in some cases
(e.g. late transition metals). 
It has been argued that this arises from
technical difficulty in including 
transitions of electrons between the $d$ and $r$ space contributing 
the screening processes. 
This oversight leads in some cases to a larger value of \emph{U}
\cite{aryasetiawan06}. Recent extensions of the cLDA method may be found
in Refs.\cite{cococcioni05} and \cite{nakamura06}. Another method for determining
the effective interaction is a scheme based on the random phase approximation
(RPA). Early attempts along this direction can be found in 
Refs.\cite{springer98,kotani00}. A combined cLDA and RPA method to 
circumvent the difficulty was also proposed \cite{solovyev05}.

Some years ago a scheme for calculating the Hubbard \emph{U}, called the
constrained RPA (cRPA) scheme \cite{aryasetiawan04}, was proposed. The main
merit of the cRPA method over currently available methods is that it allows
for a precise elimination of screening channels, which are instead to be included in
a more sophisticated treatment of the model Hamiltonian. 
This is a controlled approximation without any ambiguity, expected 
to become asymptotically exact if the $r$-space becomes well 
separated from the $d$ space.
Moreover, the effective screened interaction can be
calculated as a function of $\mathbf{r}$ and $\mathbf{r}^{\prime}$, i.e.,
$U(\mathbf{r,r}^{\prime})$, independent of the basis functions. This allows
easy access to obtaining not only on-site matrix elements but also off-site
matrix elements as well as screened exchange matrix elements, which are
usually taken to be the atomic value. Another merit is the possibility of
obtaining the frequency-dependent Hubbard \emph{U}, which may prove to be
important. The cRPA method has now been applied to a number of systems with success 
\cite{aryasetiawan06,miyake08a,nakamura08,miyake08b}.

Although the cRPA method is rather general, its applications to real systems
have revealed a serious technical problem. The problem arises when the narrow
band is entangled with other bands, i.e., it is not completely isolated from
the rest of the bands. In many materials, 
the narrow band of interest is entangled. Even in simple materials such as the 3$d$
transition metals, the 3$d$ bands mix with the 4$s$ and 4$p$ bands. Similarly, the 4$f$
bands of the 4$f$ metals hybridize with the more extended $s$ and $p$ bands. For
such cases, it is not clear anymore which part of the polarization should be eliminated
when calculating the Hubbard $U$ using the cRPA method.

Some procedures to overcome the problem of determining \emph{U} for entangled
bands have been attempted. 
One of these is to choose a set of band indices and 
define the bands of Hubbard model as those bands corresponding to the chosen indices.
Another alternative is to introduce an energy window and define the Hubbard
bands to be those that have energy within the energy window. 
Yet another alternative is to have a combination of energy window and band indices. 
These procedures, however, suffer from a number of difficulties. 
When choosing band
indices it is inevitable that some of the states will have a 
character very different from that of the intended model. For example, in the case of 3$d$
transition metals, choosing five ``3$d$" bands will include at some $\mathbf{k}%
$-points states which have little 3$d$ character, with a considerable  4$s$ component instead. 
Moreover, the chosen bands
will be awkward to model since they do not form smoothly connected bands. A
similar problem is encountered when choosing an energy window. 
A combination of band indices and energy window proposed in Ref. \cite{aryasetiawan06} 
partially solves the problem but it suffers from arbitrariness. 
Another procedure is, as we will discuss in detail later, to project the
polarization to the orbitals of interest, e.g., 3$d$ orbitals, but this procedure
has been found to yield an unphysical result of negative static \emph{U}. 
In this work, we offer a solution to the problem of determining the Hubbard
\emph{U} for entangled bands. 
The basic idea is to disentangle the narrow bands of interest from the rest 
and carry out the cRPA calculation for the disentangled band structure, 
not using the original band structure. 
The disentangling procedure is described in Sec.\ref{sec:method}. 
We apply the method to 3$d$ transition metals in Sec.\ref{sec:result} 
and show that the method is numerically stable and yields 
reasonable values of $U$. 
Finally the paper is summarized in Sec.\ref{sec:summary}.

\section{Method
\label{sec:method}
}
In the cRPA method we first choose a one-particle subspace 
$\left\{  \psi_{d}\right\}  $, which defines the model Hamiltonian, and label the rest of
the Hilbert space by $\left\{  \psi_{r}\right\}  $. We define $P_{d}$ to be
the polarization within the $d$ subspace and the total polarization is written
as $P$. It is important to realize that the rest of the polarization
$P_{r} = P - P_{d}$ is \emph{not}\textbf{ }the same as the polarization of the $r$
subspace because it contains polarization arising from transitions between the
$d$ and $r$ subspaces. 
Since $P_{d}$ is the polarization of the model
Hamiltonian, this polarization should be subtracted out from the total polarization 
when the effective parameter of the model is determined. 
The effective Coulomb interaction $W_{r}$ should
be calculated with the rest of the polarization $P_{r}$:%

\begin{equation}
W_{r}(\omega)=[1-vP_{r}(\omega)]^{-1}v\label{Wr} \;,%
\end{equation}
where $v$ is the bare Coulomb interaction.
It can indeed be shown \cite{aryasetiawan04}
that the fully screened interaction is given by%

\begin{equation}
W=[1-W_{r}P_{d}]^{-1}W_{r} \;.
\end{equation}
This mathematical identity ensures that $W_{r}$ can be interpreted as the
effective interaction among the electrons residing in the $d$ subspace since
the screening of $W_{r}$ by $P_{d}$ leads to the fully screened interaction. 
The matrix elements of $W_{r}$ in some localized
functions can then be regarded as the frequency-dependent Hubbard \emph{U}.
It has been shown that the formula in Eq.(\ref{Wr}) is formally exact,
provided $P_{r}$ is the difference between the exact polarization $P$ and the
exact polarization of the $d$ subspace $P_{d}$. In the cRPA\ method, $P_{r}$
is calculated within the random-phase approximation.

If the $d$ subspace forms an isolated set of bands, as for example in the case
of the  $t_{2g}$ bands in SrVO$_{3}$, the cRPA method can be straightforwardly
applied. However, in practical applications, the $d$ subspace may not always
be readily identified. An example of these is provided by the 3$d$ transition
metal series where the 3$d$ bands, which are usually taken to be the $d$
subspace, do not form an isolated set of bands but rather they are entangled
with the 4$s$ and 4$p$ bands. 
To handle these cases we propose the following procedure.

We first construct a set of localized Wannier orbitals from a given set of
bands defined within a certain energy window. These Wannier orbitals may be
generated by following the post-processing procedure of Souza, Marzari and Vanderbilt 
\cite{souza01,marzari97} 
or other methods, such as the preprocessing scheme proposed by Andersen {\it et al.} 
within the Nth-order muffin-tin orbital (NMTO) method \cite{andersen00}. 
We then choose this set of Wannier
orbitals as the generators of the $d$ subspace and use them as a basis for diagonalizing the
one-particle Hamiltonian, which is usually the Kohn-Sham Hamiltonian 
in the local density approximation (LDA) or generalized gradient approximation (GGA). 
The so obtained set of bands, which equivalently define the
$d$ subspace, may be slightly different from the original bands defined
within the chosen energy window, 
because hybridization effects between the $d$ and $r$ spaces are neglected. 
However, it is important to confirm that the dispersions near the 
Fermi level well reproduces the original Kohn-Sham bands. 
From these bands we calculate the polarization $\tilde{P}_{d}$ as%

\begin{equation}
\tilde{P}_{d}(\mathbf{r,r}^{\prime};\omega)=\sum_{i}^{\text{occ}}%
\sum_{j}^{\text{unocc}}\left[  \frac{\tilde{\psi}_{i}^{\ast}%
(\mathbf{r)}\tilde{\psi}_{j}(\mathbf{r)}\tilde{\psi}_{j}^{\ast}(\mathbf{r}%
^{\prime})\tilde{\psi}_{i}(\mathbf{r}^{\prime})}{\omega-\tilde{\varepsilon
}_{j}+\tilde{\varepsilon}_{i}+i\eta}-\frac{\tilde{\psi}_{i}(\mathbf{r)}%
\tilde{\psi}_{j}^{\ast}(\mathbf{r)}\tilde{\psi}_{j}(\mathbf{r}^{\prime}%
)\tilde{\psi}_{i}^{\ast}(\mathbf{r}^{\prime})}{\omega+\tilde{\varepsilon}%
_{j}-\tilde{\varepsilon}_{i}-i\eta}\right]
\;,
\label{eq:pd}
\end{equation}
where $\{  \tilde{\psi}_{i} \}$, $\{ \tilde{\varepsilon}_{i} \}  $  $(i = 1, \cdots N_d)$ are the
wavefunctions and eigenvalues obtained from diagonalizing the one-particle
Hamiltonian in the Wannier basis. 

It would seem sensible to define the rest of
the polarization as $P_{r}=P-\tilde{P}_{d},$ where $P$ is the full
polarization calculated using the \emph{original} (Kohn-Sham) wavefunctions 
and eigenvalues $\{  \psi_{i} \}$, $\{ \varepsilon_{i} \}$ $(i = 1, \cdots N)$, and calculate
$W_{r}$ according to Eq.(\ref{Wr}). We have found, however, that this procedure
is numerically very unstable, resulting in some cases to unphysically negative
static $U$ and a large oscillation as a function of frequency. 
This is understandable given that 
$P$ and $\tilde{P}_d$ are obtained from two different band structures, 
so that low energy screening channels associated with the $d$-$d$ transitions
are not excluded from $P_r$ completely. 
Due to the singular nature of the expression in Eq.(\ref{Wr}) 
these low-energy excitations can cause a large fluctuation in $W_{r}$.

Another way of calculating $P_r$ is to project the wavefunctions to the $d$ space,  

\begin{equation}
\vert \bar{\psi}_{i} \rangle = {\hat{\cal P}}
\vert \psi_i \rangle
\;,
\end{equation}
where the projection operator ${\hat{\cal P}}$ is defined as 
\begin{equation}
{\hat{\cal P}} = \sum_{j=1}^{N_d} \vert \tilde{\psi}_j \rangle \langle \tilde{\psi}_j \vert
\;.
\end{equation}

The effective $d$ polarization may be expressed as 
\begin{equation}
\bar{P}_{d}(\mathbf{r,r}^{\prime};\omega)=\sum_{i}^{\text{occ}}%
\sum_{j}^{\text{unocc}}\left[  \frac{\bar{\psi}_{i}^{\ast}%
(\mathbf{r)}\bar{\psi}_{j}(\mathbf{r)}\bar{\psi}_{j}^{\ast}(\mathbf{r}%
^{\prime})\bar{\psi}_{i}(\mathbf{r}^{\prime})}{\omega-{\varepsilon
}_{j}+{\varepsilon}_{i}+i\eta}-\frac{\bar{\psi}_{i}(\mathbf{r)}%
\bar{\psi}_{j}^{\ast}(\mathbf{r)}\bar{\psi}_{j}(\mathbf{r}^{\prime}%
)\bar{\psi}_{i}^{\ast}(\mathbf{r}^{\prime})}{\omega+{\varepsilon}%
_{j}-{\varepsilon}_{i}-i\eta}\right]
\;,
\end{equation}
and $P_r = P - \bar{P}_d$ can be used to calculate $W_r$. 
We found that this procedure does not work either and is again unstable. 
This problem may be related to the fact that 
$\bar{\psi}_i$'s are not orthogonal with each other and transitions between 
the states do not correspond to single particle-hole excitations. 

Based on these observations we propose the following procedure. 
We define the $r$ subspace by%

\begin{equation}
\vert {\phi}_{i} \rangle =
(1 - {\hat{\cal P}}) \vert \psi_{i} \rangle
\end{equation}
which is orthogonal to the $d$ subspace constructed from the Wannier orbitals.
In practice it is convenient to orthonormalize $\{ \phi_i \}$ and prepare 
$N-N_d$ basis functions. 
By diagonalizing the Hamiltonian in this subspace 
a new set of wavefunctions $\{ \tilde{\phi}_i \}$ 
and eigenvalues $\{ \tilde{e}_i \}$ $(i=1,\cdots, N-N_d)$ are obtained. 
As a consequence of orthogonalizing 
$\{  \tilde{\phi}_{i} \} $  and $\{ \tilde{\psi}_{j} \} $, 
the set of $r$ bands $\{ \tilde{e}_i \}$ are completely
disentangled from those of the $d$ space $\{  \tilde{\varepsilon}_{j} \} $, 
and they are slightly different from the original band structure $\{ \varepsilon_i \}$.  
As we will see later, however, the numerical tests show that 
the disentangled band structure is close to the original one. 

The Hubbard $U$ is calculated 
according to Eq.(\ref{Wr}) with $P_{r}=\tilde{P}-\tilde{P}_{d}$, where 
$\tilde{P}$ is the full polarization calculated for the {\it disentangled} 
band structure. 
It is important to realize that the screening processes 
between the $d$ space and the $r$ space are included in $U$, 
although the $d$-$r$ coupling is cut off in the construction of 
the wavefunctions and eigenvalues.

\section{Results and discussions 
\label{sec:result}} 

As an illustration we apply the method to 3$d$ transition metals. 
The electronic structure calculations are done in the local density approximation \cite{kohn65}
of density functional theory \cite{hohenberg64} 
with the full-potential LMTO implementation \cite{methfessel00}. 
The wavefunctions are expanded by $spdf+spd$ MTOs 
and a $8\times8\times8$ {\bf k}-mesh is used for 
the Brillouin zone summation. Spin polarization is neglected. 
More technical details are found elsewhere 
\cite{schilfgaarde06,miyake08a}. 

\begin{figure}
\begin{center}
\includegraphics[width=90mm]{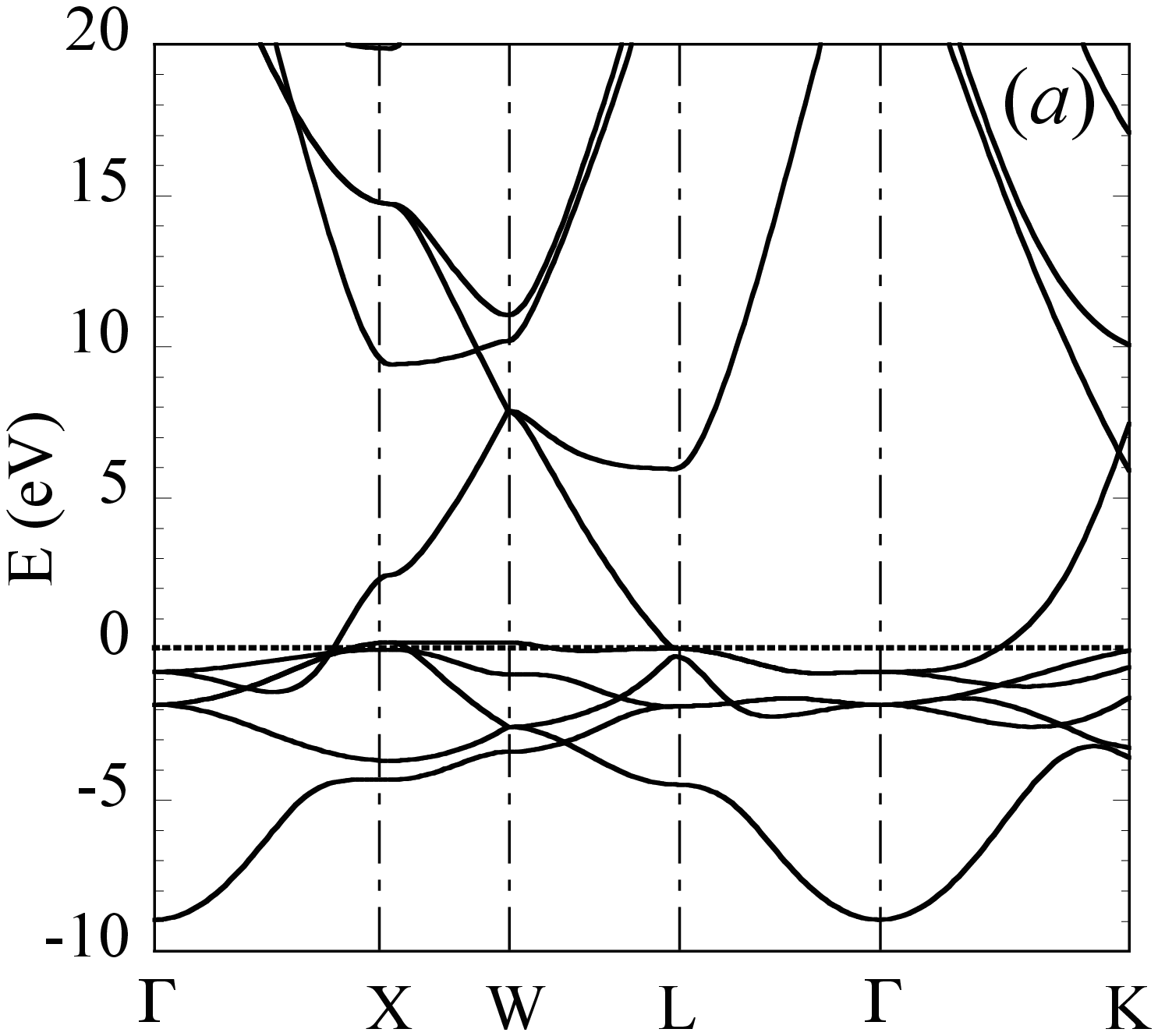}
\includegraphics[width=90mm]{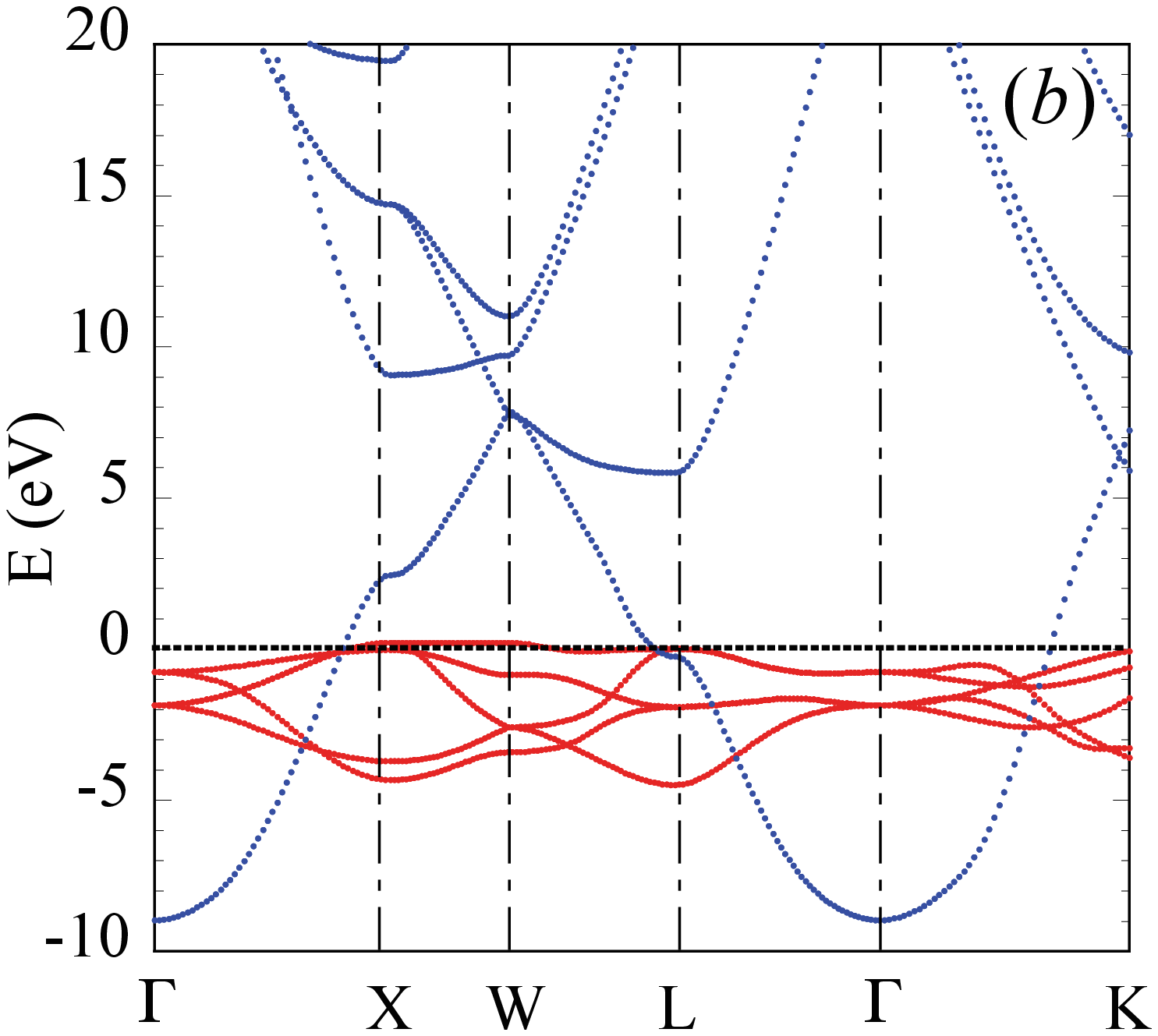}
\end{center}
\caption{(Color)
(a) Kohn-Sham band structure of nickel in the LDA.
(b) Disentangled band structure with $d$-$r$ hybridization switched off. 
The red lines show the $d$ states obtained by 
the maximally localized Wannier scheme, 
while the blue lines are disentangled $r$ states. 
Energy is measured from the Fermi level.
}%
\label{fig:band}%
\end{figure}

Figure \ref{fig:band}(a) shows the Kohn-Sham band structure of nickel. 
There are five orbitals having strong 3$d$ character at [-5 eV:1 eV], 
crossed by a dispersive state which is of mainly 4$s$ character. 
Using the maximally localized Wannier function prescription 
with the energy window of [-7 eV:3 eV], 
interpolated ``$d$'' bands are obtained. 
The subsequent disentangling procedure gives the associated ``$r$'' bands. 
Comparing Fig.\ref{fig:band}(b) with (a) we can see that 
there is no anti-crossing between the $d$ bands and the $r$ bands in (b).
Otherwise the two band structures are nearly identical. 

\begin{figure}
\begin{center}
\includegraphics[width=90mm]{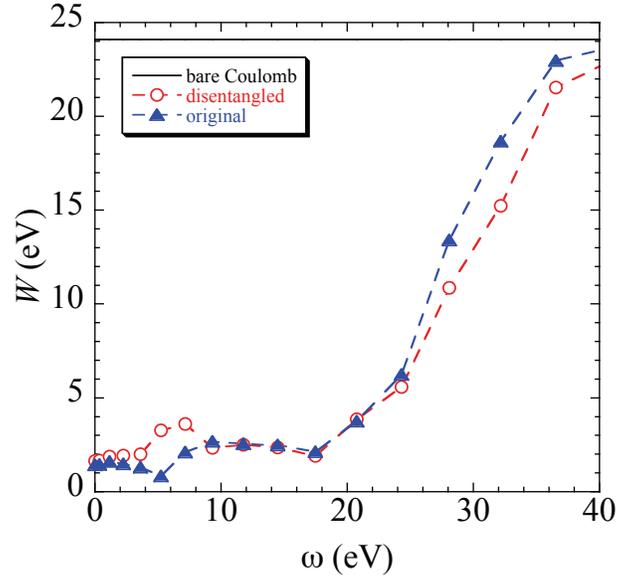}
\end{center}
\caption{(Color online)
Fully screened Coulomb interaction of nickel as a function of frequency.
The average of the diagonal terms in the Wannier basis is plotted. 
The present scheme using the disentangled bands is compared to 
the results from the original band structure.
}%
\label{fig:w}%
\end{figure}

In order to see the impact of the disentanglement on the screening effects, 
we perform the full RPA calculation using  
the disentangled band structure. 
The fully screened Coulomb interaction is compared 
with that for the original band structure  
in Fig.\ref{fig:w} where the average of the five diagonal terms 
in the Wannier basis $\varphi_i$ is plotted, 

\begin{equation}
W(\omega) = \frac{1}{5}\sum_{i=1}^{5}
\int d {\bf r} d{\bf r'} \varphi_i^*({\bf r}) \varphi_i({\bf r}) 
W({\bf r},{\bf r'};\omega)
\varphi_i^*({\bf r'}) \varphi_i({\bf r'})
\;.
\label{eq:W}
\end{equation}
The two methods yield similar results 
The static values agree with each other within 0.2 eV, and 
the frequency dependence is weak at low frequencies. 
As frequency increases there is a sharp increase at $\sim$20 eV, 
where screening by plasmons becomes ineffective. 
These results assure that screening effects can be treated 
accurately with the disentangled band structure. 

\begin{figure}
\begin{center}
\includegraphics[width=90mm]{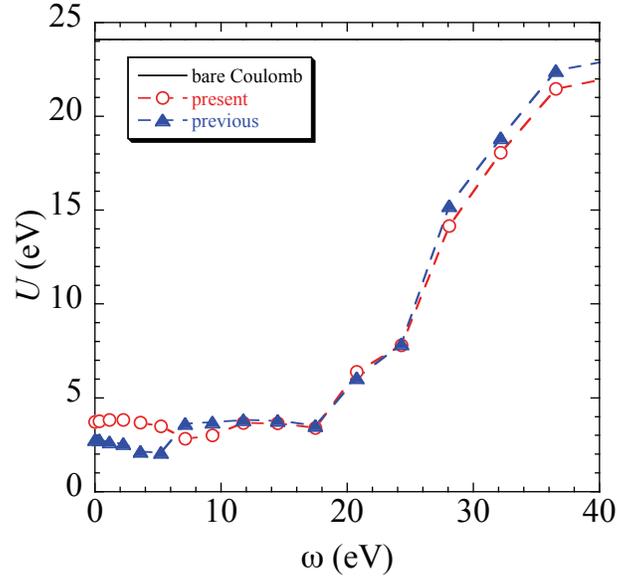}
\end{center}
\caption{(Color online)
Hubbard $U$ of nickel as a function of frequency. 
The diagonal term of the partially screened interaction 
in the Wannier basis is calculated 
by the present method and compared with the 
published data of Ref.\cite{miyake08a}. 
}
\label{fig:u}%
\end{figure}

The Hubbard $U$ is calculated by the constrained RPA, namely, 
by replacing $W$ in Eq.(\ref{eq:W}) with $W_r$. 
The results are shown in Fig.\ref{fig:u}. 
There is no large fluctuation against frequency, 
in contrast to the methods described in Sec.\ref{sec:method}, 
and $U(\omega)$ shows a stable behavior. 
As is expected, $U$ is significantly larger than $W$ at low frequencies. 
This implies proper elimination of $d-d$ screening processes is crucial. 
Comparing with the previous results using a combined energy and band window
\cite{miyake08a}, the agreement is reasonably good. 
A small difference between these two results at low frequency may be due to 
a small portion of $d$-$d$ screening presumably contained in the 
previous method \cite{miyake08a}, although it should be excluded from the 
cRPA calculations. 

\begin{figure}
\begin{center}
\includegraphics[width=90mm]{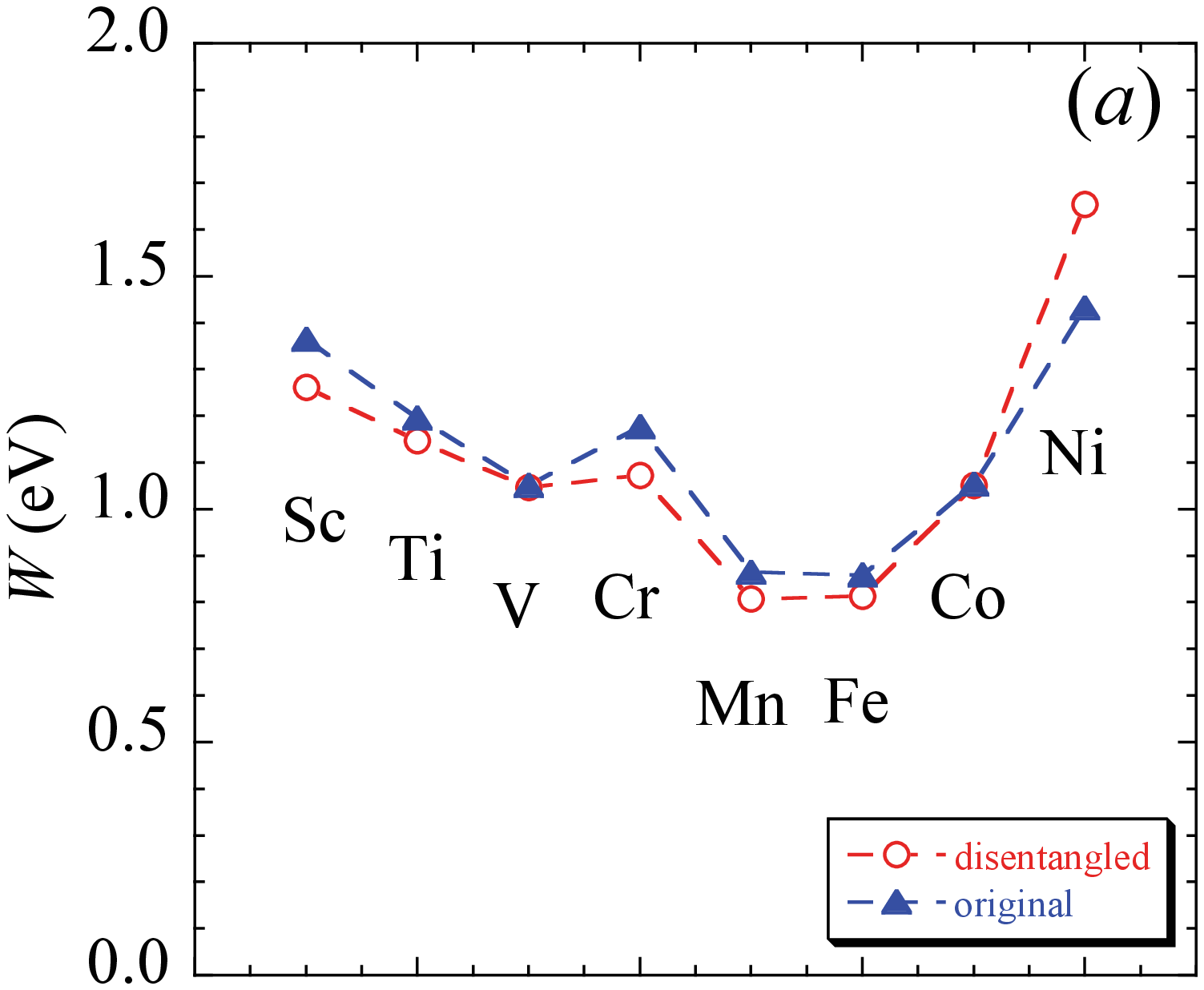}
\includegraphics[width=90mm]{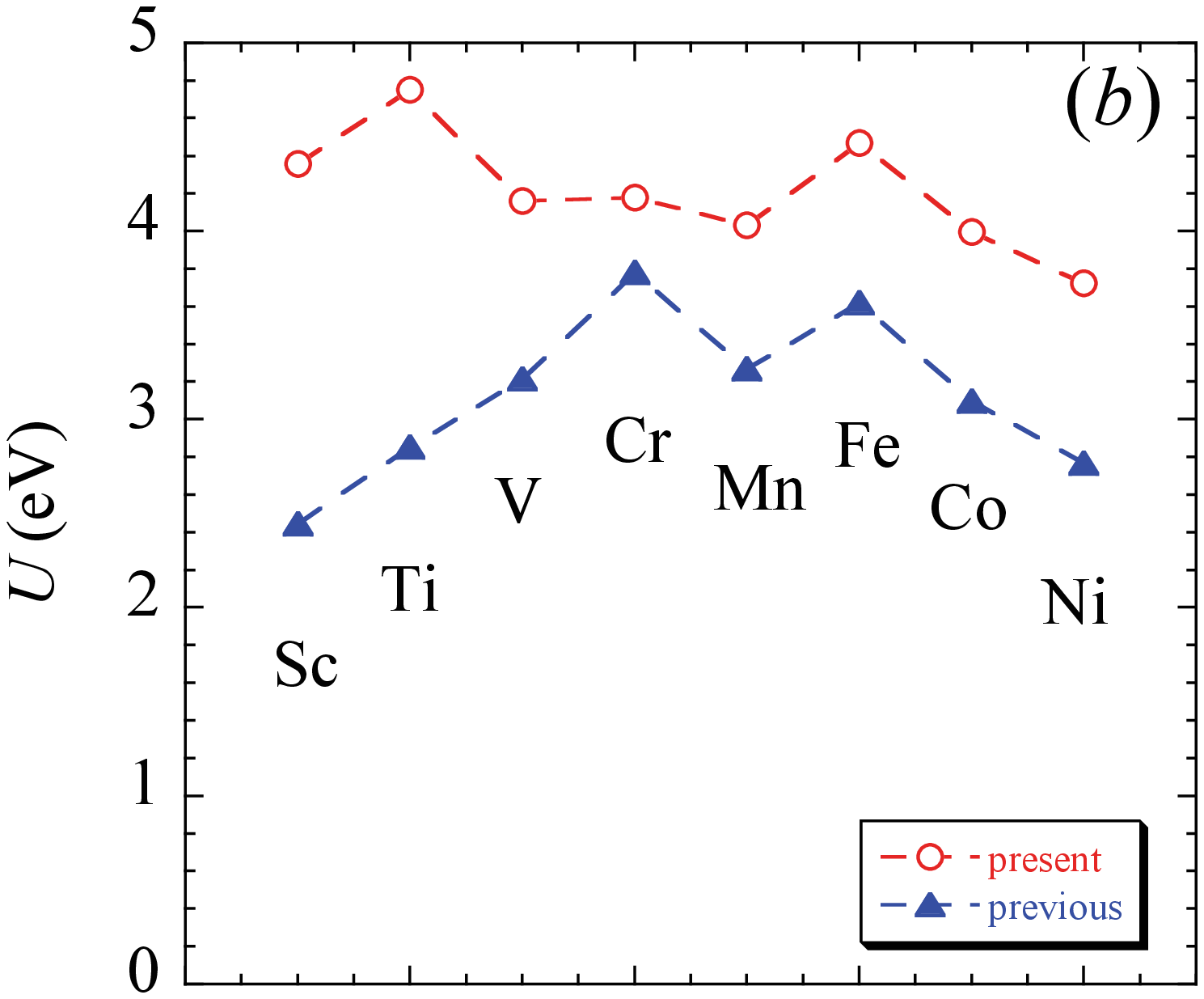}
\end{center}
\caption{(Color online)
Static values of (a) fully screened Coulomb interaction $W$ 
and (b) Hubbard $U$ for 3$d$ metals.
}%
\label{fig:static}%
\end{figure}

We carried out the calculations for a series of other 3$d$ metals as well and found 
in all the cases that 
(i) the present scheme is numerically stable and does not result in 
unphysical frequency dependence of $U$, and 
(ii) the value of $W(\omega)$ is close to that from the original band structure. 
The latter is confirmed in Fig.\ref{fig:static} where the static values 
($\omega\rightarrow 0$ limit)
are summarized. 
Concerning $U$, 
the present method gives larger values compared to the 
previous results, particularly for early transition metals. 
Since $W$ is nearly equal to each other in the two methods, 
the discrepancy is ascribed to the different treatment 
between $P_d$ and ${\tilde P}_d$. 
We should note that $P_d$ in the previous method depends on 
the choice of the window. For a wider window, obviously we would obtain 
a better agreement. 
Also, some states have a mixed character of 3$d$ and 4$s$ 
near the anti-crossing points. 
This makes elimination of the screening process difficult in 
the original band structure.
The present scheme, on the other hand, enables us to determine 
$U$ without ambiguity. 
The $d$ bands are disentangled from the $r$ bands. 
Consequently, the polarization in the $d$ space is well-defined and 
can be removed completely in $P_r = \tilde{P} - {\tilde P}_d$. 

In the present formulation, small off-diagonal matrix elements of the Kohn-Sham Hamiltonian between 
the $d$ space wavefunction $\vert \psi_i\rangle $ constructed from the Wannier orbitals and 
the $r$ space $\vert \phi_j\rangle $ are ignored. This is the reason why the anti-crossing is avoided.  
If the energy of this hybridization point in the band dispersion is smaller than the screened Coulomb 
energy and the energy scale of the interest, strictly speaking, one has to keep all of these hybridizing 
bands in the effective model, because the hybridization effects are non-perturbative. In the present 
case of transition metals, the energy crossing point of 4$s$ and 3$d$ bands are relatively larger 
than the screened Coulomb energy scale and the low energy models constructed only from 
the 3$d$ Wannier orbitals may give at least qualitatively reasonable description of 
the low energy physics.    
%%%%%%%%%%%%

\section{Summary
\label{sec:summary}}
We have proposed a method to calculate the effective interaction parameters 
for the effective low-energy models of real materials when bands are entangled. 
The key point is 
to first properly orthogonalize the low-energy subspace contained in the models and 
the complementary high-energy subspace to each other. 
This orthogonalization by the projection technique enables the disentanglement 
of the bands. 
Once the disentangled band structure is obtained, 
the constraint RPA method can be used to determine the partially screened 
Coulomb interaction uniquely. 
Numerical tests for 3$d$ metals show that the method is stable 
and yields reasonable results. 
The method is applicable to any system. 
Applications to more complicated systems, such as interfaces of 
transition metal oxides are now under way.

\acknowledgements 
We thank S. Biermann for fruitful discussions. 
This work was supported by Grants-in-Aid for Scientific Research from MEXT, Japan 
under grant numbers 17064004 and 19019013. 
TM acknowledges support from 
Next Generation Supercomputing Project, Nanoscience Program, MEXT, Japan. 
\newline

%\begin{thebibliography}{99}                                                                                               %

\newpage

\end{document}